\documentclass[11pt]{article}
\usepackage{amsmath,amssymb,epsfig,bbold, cite,amscd}
\usepackage {color}
\advance\hoffset by -1.1truecm
\advance\voffset by -1.0truecm
\newcommand{\be}{\begin{equation}}
\newcommand{\ee}{\end{equation}}
\newcommand{\bea}{\begin{eqnarray}}
\newcommand{\eea}{\end{eqnarray}}

\usepackage[numbers,sort&compress]{natbib}
\usepackage[colorlinks=true
,urlcolor=blue
,anchorcolor=blue
,citecolor=blue
,filecolor=blue
,linkcolor=blue
,menucolor=blue
,pagecolor=blue
,linktocpage=true
,pdfproducer=medialab
,pdfa=true
]{hyperref}

\numberwithin{equation}{section}

\newcounter{appendice}

\begin{document}

 \title{
\vspace{2.2cm}\begin{flushleft} \bf {Fuzzy Conifold $Y_F^6$ and Monopoles on $S_F^2\times S_F^2$ 
\linethickness{.05cm}\line(1,0){433}
}\end{flushleft}}
\author{\bf{Nirmalendu Acharyya\footnote{nirmalendu@cts.iisc.ernet.in},\, and Sachindeo Vaidya\footnote{vaidya@cts.iisc.ernet.in}} \\ 
\begin{small}{\it Centre for High Energy Physics, Indian Institute of Science, Bangalore 560012, India}
\end{small} \\
}
\date{\empty}
\maketitle
\abstract
In this article, we construct the fuzzy (finite dimensional) analogues of the conifold $Y^6$ and its base $X^5$. 
We show that fuzzy $X^5$ is (the analogue of) a principal $U(1)$ bundle over fuzzy spheres $S^2_F \times S^2_F$ and 
explicitly construct the associated monopole bundles. In particular our construction provides an explicit 
discretization of the spaces $T^{\kappa,\kappa}$ and $T^{\kappa,0}$.
\newpage
\hspace{-1cm}\linethickness{.01cm}\line(1,0){430}
\tableofcontents
\hspace{-1cm}\linethickness{.01cm}\line(1,0){430}

\section{Introduction}

The conifold $Y^{2n-2}$ is a  $(2n-2)$ dimensional manifold embedded in a $n$-dimensional complex plane $\mathbb{C}^n$, with cone-like singularities at isolated point(s). In the neighbourhood of such a singularity, a conifold can be described by a quadratic in $\mathbb{C}^n$  \cite{Candelas:1989js}:
\begin{equation}
\sum_{\alpha=1}^n z_\alpha^2 =0.
\label{cone_ndim}
\end{equation}
Here, the singular point is chosen to be at the origin of $\mathbb{C}^n$. $Y^{2n-2}$ is smooth everywhere except at $z_\alpha =0$, which is the apex of the cone.

In matrix notation, (\ref{cone_ndim}) can be written as 
\begin{equation}
z^T z =0, \quad\quad\quad z=\left(\begin{array}{lll}
z_1\\z_2\\z_3\\\vdots \\z_n
\end{array}\right).
\end{equation}
 Under the transformation 
\begin{equation}
z \rightarrow \mathcal{R}z, \quad\quad \mathcal{R}^T \mathcal{R}=1, \quad \quad \mathcal{R}\in SO(n),
\end{equation}
the condition (\ref{cone_ndim}) is invariant. So $Y^{2n-2}$ is $SO(n)$ symmetric. 
It  also admits an additional U(1) symmetry  $z\rightarrow e^{i \lambda}z, \,\,\, \lambda \in \mathbb{R}$.
Thus the symmetry group of this space is $SO(n) \times U(1)$.

The conifold $Y^{2n-2}$ has a $(2n-3)$-dimensional compact manifold $X^{2n-3}$ as its base.
$X^{2n-3}$ is the intersection of  $Y^{2n-2}$ with $S^{2n-1}$,  the latter  defined by 
\begin{equation}
z^\dagger z=\textrm{fixed}.
\end{equation} 
$Y^{2n-2}$ is a cone over $X^{2n-3}$. $X^{2n-3}$ is a compact Einstein manifold: $\left(R_{ab}\right)_{X^{2n-3}} = (2n-4)\left(g_{ab}\right)_{X^{2n-3}}$ while $Y^{2n-2}$ is Ricci flat \cite{Klebanov:1998hh}.

Here we are interested in the $n=4$ case.
This space is important in the context of  gauge-gravity duality 
(see \cite{Klebanov:1998hh, Pando Zayas:2000sq, Maldacena:1997re,Gubser:1998ia,Dasgupta:1998su}).

For  $Y^6$, the symmetry group is $SO(4)\times U(1)\simeq SU(2)\times SU(2)\times U(1)$. 
The base $X^5$ has the topology of $ S^3 \times S^2$ and is in fact a $U(1)$ bundle over $S^2\times S^2$. 
 $X^5$ belongs to the class of manifolds $T^{p,q}$ which have the topology of  $S^3 \times S^2$, but different $T^{p,q}$'s have
 different geometries.

In section \ref{sec2} we review the conifold $Y^6$ and its base $X^5$. We recall their geometric properties and the symmetries, especially those useful
for our subsequent discussion.

In  section \ref{sec3} we show that $X^5$ is a principal $U(1)$ bundle over $S^2\times S^2$.
 A ``Hopf-like"  map  from $X^5 \rightarrow S^2\times S^2$ can be written using the 4--dimensional representation of $SU(2) \times SU(2)$. 
This $U(1)$ bundle is nontrivial and corresponds to a magnetic monopole, while the associated line bundles describe nontrivial  
configurations of a complex scalar field on $S^2 \times S^2$.
 In section \ref{sect_com_line_bundle} we show that  these complex scalar fields can be naturally expanded in terms of the functions of 
 $SU(2)\times SU(2)$.
This is very useful as it makes subsequent {\it fuzzification} straightforward.

In section \ref{kaluza_klein_sec}, we present this monopole as a Kaluza-Klein monopole in the spirit of 
\cite{Sorkin:1983ns, Gross:1983hb, RandjbarDaemi:1982hi}. The fact that the total space and the base space are both Einstein manifolds provides us with 
an interesting bound for the action of the gauge field.

Our intention is to extend this construction to fuzzy spaces.
To write the fuzzy analogue, in section \ref{fuzzy_section} we start by describing the fuzzy versions of the $Y^6$ and $X^5$ 
and then specify a Jordan-Schwinger map $X^5_F \rightarrow S_F^2\times S_F^2$. The  fuzzy spheres $S_F^2\times S_F^2$ are 
finite dimensional representations of $SU(2) \times SU(2)$ and $X^5_F$ is a noncommutative bundle over it.
In section \ref{fuzzy_fibre_bundle} we  describe the fuzzification of   section \ref{sect_com_line_bundle}  
(following \cite{Grosse:1995jt, Acharyya:2013hga}) to identify the associated complex line bundle. We end this section 
by illustrating our construction with simple examples.

Section \ref{tpq} deals with the spaces $T^{p,q}$ and their fuzzy analogues. After a brief review of continuum $T^{p,q}$ and its description as Hopf fibration, 
we obtain the monopole by Kaluza-Klein reduction. The fuzzification is again straightforward, which we show in section \ref{tpq1}.

\section{ The Conifold $Y^6$ and Its Base $X^5$} \label{sec2}

The conifold $Y^6$  is a 6--dimensional manifold  embedded in $\mathbb{C}^4$ with 4 complex coordinates $z_\alpha$  $(\alpha=1,...4)$ satisfying \cite{Candelas:1989js}
\begin{equation}
\mathcal{O}_4(z_\alpha)\equiv z_1 z_4-z_2 z_3 =0, \quad z_\alpha \in \mathbb{C}^4.
\label{conifold_1}
\end{equation}
It is a   smooth manifold with conical singularity  at a single point $z_\alpha=0 $ where the function $\mathcal{O}_0(z_\alpha)$ and its  derivatives vanish:
\begin{equation}
\mathcal{O}_4(z_\alpha)|_{z_\alpha=0} =0, \quad \left(\frac{\partial\mathcal{O}_4}{\partial z_\alpha}\right)_{z_\alpha=0} =0.
\end{equation}
This complex manifold $Y^6$ is the set of all lines  in $\mathbb{C}^4$ passing through origin and hence  a cone  
with the double singular point $z_\alpha=0$ as its  apex.

The intersection of $Y^6$ with $S^7\in\mathbb{C}^4$ ($z^\dagger z =R^2$) is called $X^5$, the base of the conifold. For $R=$constant $ > 0$, $X^5$ is  void of any singularity and   is  a smooth 5--dimensional manifold. 
It can be described in terms of   four complex numbers obeying
\begin{equation}
\mathcal{O}_4\equiv z_1 z_4-z_2z_3=0, \quad   \bar{z}_1z_1+\bar{z}_2 z_2+\bar{z}_3z_3+\bar{z}_4z_4=R^2.
\label{deformed_conifold}
\end{equation}

This manifold $X^5$  can be parametrized by five angular coordinates \cite{ {Pando Zayas:2000sq}} : 
\begin{eqnarray}
&&z_1=  R e^{\frac{i}{2}(\phi+\zeta_1+\zeta_2)} \cos\frac{\theta_1}{2}\cos\frac{\theta_2}{2}, \quad z_2=R e^{\frac{i}{2}(\phi+\zeta_1-\zeta_2)} \cos\frac{\theta_1}{2}\sin\frac{\theta_2}{2}, \nonumber \\
&&z_3= R e^{\frac{i}{2}(\phi-\zeta_1+\zeta_2)} \sin\frac{\theta_1}{2}\cos\frac{\theta_2}{2}, \quad z_4 =  R e^{\frac{i}{2}(\phi-\zeta_1-\zeta_2)} \sin\frac{\theta_1}{2}\sin\frac{\theta_2}{2}
\label{parametrization}
\end{eqnarray}
with  $0\leq\theta_a<\pi$, \,\, $0\leq \zeta_a <2\pi$ and $0\leq \phi<4\pi$.

The metric on  $X^5$ can be written in terms of these angular coordinates \cite{Candelas:1989js, Page:1984ae, Klebanov:1998hh, Ohta:1999we}:
\begin{equation}
ds^2_{X^5}=\frac{\rho^2}{9}{\Big(}d\phi+  \cos \theta_1 d \zeta_1 + \cos \theta_2 d \zeta_2{\Big)}^2+ \frac{\rho^2}{6}\sum_{a=1}^2 {\Big(}d\theta_a^2 + \sin ^2 \theta_a d\zeta_a^2{\Big)}
\label{metric}
\end{equation}
where $\rho^2 = \frac{3}{2} R^{4/3}$. This metric  satisfies the Einstein condition 
\begin{equation}
\tilde{R}_{ab}= \frac{4}{\rho^2} \tilde{g}_{ab}.
\end{equation}
The metric on the $Y^6$ 
\begin{equation}
ds^2_{Y^6} = d\rho^2 + ds^2_{X^5}
\end{equation}
 is Ricci--flat.

There is a scaling symmetry on $Y^6$.
Under $z_\alpha \rightarrow \beta z_\alpha, \,\,\, \beta \in \mathbb{C}$, (\ref{conifold_1}) is invariant. But the metric gets rescaled: $ds_{Y^6}^2 \rightarrow |\beta|^2 ds_{Y^6}^2$.

In order to facilitate the discussion on fibre bundles (see section \ref{kaluza_klein_sec}), we do the following coordinate transformation:
\begin{eqnarray}
&&\zeta_a\rightarrow \xi_a = 2 \pi - \zeta_a , \nonumber\\
&&\phi \rightarrow \psi = \frac{1}{2} \left(\phi - \xi_1 -\xi_2\right).
\end{eqnarray}
 The range of $\psi$ is $[0,4\pi)$.
Under $\phi \rightarrow \psi, \,\,\, \zeta_a \rightarrow \xi_a $, the parametrization (\ref{parametrization}) becomes 
\begin{eqnarray}
&&z_1=  R e^{i\psi} \cos\frac{\theta_1}{2}\cos\frac{\theta_2}{2}, \quad   \,\,\, \quad z_2=R e^{i(\psi+\xi_2)} \cos\frac{\theta_1}{2}\sin\frac{\theta_2}{2}, \nonumber \\
&&z_3= R e^{i(\psi+\xi_1)} \sin\frac{\theta_1}{2}\cos\frac{\theta_2}{2}, \quad z_4 =  R e^{i(\psi+\xi_1+\xi_2)} \sin\frac{\theta_1}{2}\sin\frac{\theta_2}{2}
\label{parametrization2}
\end{eqnarray}
and the metric on $X^5$ can be rewritten as
\begin{equation}
ds^2_{X^5}=\frac{\rho^2}{9}{\Big(}2d\psi +(1- \cos \theta_1) d \xi_1 +(1- \cos \theta_2) d \xi_2{\Big)}^2+ \frac{\rho^2}{6}\sum_{a=1}^2 {\Big(}d\theta_a^2 + \sin ^2 \theta_a d\xi_a^2{\Big)}.
\label{metric2}
\end{equation}
If we choose the $S^5$ to be of unit radius ($R=1$), then $\rho^2 =\frac{3}{2} $ and 
\begin{equation}
ds^2_{X^5}=\frac{2}{3}{\Big(}d\psi +\sin^2 \frac{\theta_1}{2} d \xi_1 +\sin^2 \frac{\theta_2}{2} d \xi_2{\Big)}^2+ \frac{1}{4}\sum_{a=1}^2 {\Big(}d\theta_a^2 + \sin ^2 \theta_a d\xi_a^2{\Big)}.
\label{metric3}
\end{equation}

\section{$X^5$ is a $U(1)$ Bundle over $S^2 \times S^2$ }\label{sec3}

By exploiting the symmetries of $X^5$, we can define a map analogous to the Hopf map $S^3 \rightarrow S^2$. Let us first briefly recall the standard Hopf map. Embedded in the two--dimensional complex plane $\mathbb{C}^2$,  $S^3$ is  defined by the constraint 
\begin{equation}
\bar{z}_1 z_1 +\bar{z}_2 z_2 = \mathrm{constant}
\end{equation}
and has $SU(2)$ as the symmetry group. Using the Pauli matrices $\sigma_i $, the Hopf map is defined as 
\begin{equation}
x_i = \frac{1}{2}z_a (\sigma_i)_{ab} z_b, \quad \mathrm{where} \,\, a,b=1,2 \,\, \mathrm{and}\,\, i=1,2,3.
\end{equation}
For $z_a \in S^3$,  $x_i \in S^2$. Great circles $S^1$ on $S^3$ are mapped to  points on $S^2$ and the surjection has a $U(1)$ symmetry: $S^2 = S^3/U(1)$.  Hence, $S^3$ is a $U(1)$ fibre bundle over $S^2$  :
\begin{eqnarray}
U(1) \quad\longrightarrow & S^3 \nonumber \\ 
& \downarrow \nonumber\\
& S^2. \nonumber
\end{eqnarray}

A similar construction may be defined on $X^5$.
Here, $SO(4)$ is the symmetry group generated by $\{M_1, M_2, M_3, T_1, T_2,T_3\}$ satisfying the Lie algebra 
\begin{equation}
[M_i, M_j] = i\epsilon_{ijk}M_k, \quad [M_i, T_j] = i\epsilon_{ijk}T_k,\quad [T_i, T_j] = i\epsilon_{ijk}M_k,\quad \quad i=1,2,3. 
\end{equation}
In the fundamental representation, these are $4\times 4$ hermitian matrices
\begin{eqnarray}
M_1= \left(\begin{array}{cccc}
0&0&0&0\\
0&0&-i&0\\
0&i&0&0\\
0&0&0&0\\
\end{array}\right), \quad 
M_2= \left(\begin{array}{cccc}
0&0&i&0\\
0&0&0&0\\
-i&0&0&0\\
0&0&0&0\\
\end{array}\right),\quad
M_3= \left(\begin{array}{cccc}
0&-i&0&0\\
i&0&0&0\\
0&0&0&0\\
0&0&0&0\\
\end{array}\right)\nonumber\\
T_1= \left(\begin{array}{cccc}
0&0&0&-i\\
0&0&0&0\\
0&0&0&0\\
i&0&0&0\\
\end{array}\right), \quad 
T_2= \left(\begin{array}{cccc}
0&0&0&0\\
0&0&0&-i\\
0&0&0&0\\
0&i&0&0\\
\end{array}\right),\quad
T_3= \left(\begin{array}{cccc}
0&0&0&0\\
0&0&0&0\\
0&0&0&-i\\
0&0&i&0\\
\end{array}\right).
\end{eqnarray}

Particular linear combinations of these generators 
\begin{equation}
Q_i= \frac{1}{2} \left(M_i+T_i\right), \quad \bar{Q}_i= \frac{1}{2} \left(M_i-T_i\right), \quad \quad i=1,2,3
\end{equation}
decompose the Lie algebra of $SO(4)$ into two disjoint $SU(2)$ algebras
\begin{equation}
[Q_i,Q_j]=i\epsilon_{ijk}Q_k,\quad [\bar{Q}_i,\bar{Q}_j]=i\epsilon_{ijk}\bar{Q}_k,\quad [Q_i, \bar{Q}_j]=0. 
\label{so4_matrices_decomposed}
\end{equation}

In the fundamental representation, the relation of $SO(4)$ with the $SU(2)$'s becomes more apparent with the unitary transformation 
\begin{eqnarray}
A_i= U^\dagger Q_i U= \frac{1}{2}{I}_2 \otimes \sigma_i , \quad B_i = U^\dagger \bar{Q}_i U=\frac{1}{2}\sigma_i \otimes {I}_2,\nonumber\\
\mathrm{where}\quad U=\frac{1}{\sqrt{2}}\left(\begin{array}{cccc}
1&0&0&-1\\
i&0&0&i\\
0&-1&-1&0 \\
0&-i&i&0 
\end{array}\right).
\label{matrices}
\end{eqnarray}
This preserves the decomposition (\ref{so4_matrices_decomposed})
\begin{equation}
[A_i, A_j]=i\epsilon_{ijk} A_k, \quad [B_i, B_j]=i\epsilon_{ijk} B_k, \quad [A_i, B_j]=0.\nonumber
\end{equation}

With $A_i$ and $B_i$ we can define a map $\Pi:\mathbb{C}^4 \rightarrow \mathbb{R}^6$ 
\begin{equation}
\omega_i =z^\dagger A_i z, \quad y_i=z^\dagger B_i z, \quad  \mathrm{with}\quad z=\left(\begin{array}{cccc}
z_1\\z_2\\z_3\\z_4
\end{array}\right) \in \mathbb{C}^4
\label{map1}
\end{equation}
and $i=1,2,3$. $\omega_i$ and $y_i$ are real: $\bar{\omega}_i = \omega_i$,\,\,$\bar{y}_i=y_i$.

 $\omega_i, y_i\in\mathbb{R}^6$ satisfy the relation
\begin{eqnarray}
\sum_{i=1}^3 \omega_i^2= \sum_{i=1}^3 y_i^2 &=&\left(\frac{1}{2} \sum_{\alpha=1}^4 \bar{z}_\alpha z_\alpha \right)^2 -\left(\bar{z}_1\bar{z}_4-\bar{z}_2\bar{z}_3\right)\left(z_1z_4-z_2z_3\right) \nonumber \\
 &=&\frac{1}{4} R^2 -\bar{\mathcal{O}}_4\mathcal{O}_4.
\label{casimir1}
\end{eqnarray}

If $z_\alpha \in X^5$, then $\mathcal{O}_4=0$ and $R=1$ and (\ref{casimir1}) reduces to the constraints
\begin{equation}
\sum_{i=1}^3 \omega_i^2= \sum_{i=1}^3 y_i^2=\frac{1}{4}. 
\label{casimir678}
\end{equation}
This space is $S^2\times S^2$: the radius of each $S^2$ is $\frac{1}{2}$ and (\ref{map1}) is a map  $\Pi: X^5  \rightarrow  S^2\times S^2$.

As is obvious from (\ref{deformed_conifold}), $X^5$ has a $U(1)$ symmetry  $z\rightarrow e^{i\lambda} z$. The map (\ref{map1}) is also invariant under the same transformation. 
Using (\ref{parametrization2}) in  (\ref{map1}), we see that $S^2 \times S^2$ has coordinates
\begin{eqnarray}
&& \omega_1= \frac{1}{2}\sin \theta_2 \cos \xi_2, \quad \omega_2= \frac{1}{2}\sin \theta_2 \sin \xi_2,\quad \omega_3= \frac{1}{2}\cos \theta_2,\nonumber\\
&& y_1= \frac{1}{2}\sin \theta_1 \cos \xi_1, \quad y_2= \frac{1}{2}\sin \theta_1 \sin \xi_1,\quad y_3= \frac{1}{2}\cos \theta_1.
\end{eqnarray}
This parametrization is independent of the angle $\psi$.  This means (\ref{map1}) is a map of circles $e^{i\psi}$ on $X^5$ to points on $S^2\times S^2$. Thus $X^5$ is a  principal $U(1)$ bundle over  $S^2 \times S^2$, and the map $\Pi$ gives the fibration
\begin{eqnarray}\label{fibration}
U(1) \quad\longrightarrow & X^5 \nonumber \\ 
& \downarrow \\
& S^2\times S^2. \nonumber
\end{eqnarray}
The  $U(1)$ gauge symmetry  leads  to  monopoles on $S^2 \times S^2$.

\subsection{Monopoles on $S^2 \times S^2$}\label{sect_com_line_bundle}
It is useful to explicitly write down the coordinate charts that will be used on $S^2\times S^2$. The set of complex functions
\begin{eqnarray}
&& \chi^{NN}_1=   \cos\frac{\theta_1}{2}\cos\frac{\theta_2}{2}, \nonumber \\
&&\chi^{NN}_2= e^{i\xi_2} \cos\frac{\theta_1}{2}\sin\frac{\theta_2}{2}, \nonumber \\
&& \chi^{NN}_3=  e^{i\xi_1} \sin\frac{\theta_1}{2}\cos\frac{\theta_2}{2}, \nonumber \\
&& \chi^{NN}_4 =  e^{i(\xi_1+\xi_2)} \sin\frac{\theta_1}{2}\sin\frac{\theta_2}{2}
\label{coordinate_chart_N}
\end{eqnarray}
is well--defined on  $S^2 \times S^2$ except at two isolated points $\theta_1 =\pi$ and $\theta_2=\pi$, the {\it south poles} of the 2-spheres. We denote this coordinate chart as $U_{NN}$. It is easy to see that (\ref{coordinate_chart_N}) is obtained by choosing $\psi=0$ in the parametrization (\ref{parametrization2}). The coordinate chart $U_{SS}$ which includes the south poles $\theta_1=\pi$ and $\theta_2=\pi$ can be obtained by putting $\psi=-(\xi_1+\xi_2)$ in (\ref{parametrization2}):
\begin{eqnarray}
&& \chi^{SS}_1= e^{-i(\xi_1+\xi_2)}  \cos\frac{\theta_1}{2}\cos\frac{\theta_2}{2}, \nonumber \\
&&\chi^{SS}_2= e^{-i\xi_1} \cos\frac{\theta_1}{2}\sin\frac{\theta_2}{2}, \nonumber \\
&& \chi^{SS}_3=  e^{-i\xi_2} \sin\frac{\theta_1}{2}\cos\frac{\theta_2}{2}, \nonumber \\
&& \chi^{SS}_4 =   \sin\frac{\theta_1}{2}\sin\frac{\theta_2}{2}.
\label{coordinate_chart_S}
\end{eqnarray}
These functions are well-defined at all points on $S^2\times S^2$ except at the {\it north poles} ($\theta_1=0$ and $\theta_2=0$). 
On the overlapping region $U_{NN}\cap U_{SS}$,
\begin{equation}
\chi_\alpha^{NN} = e^{i(\xi_1+\xi_2)} \chi_\alpha^{SS}.
\end{equation}

The other two charts  $U_{NS}$ and $U_{SN}$ can be obtained by  putting $\psi=-\xi_1$ and $\psi=-\xi_2$  respectively in  (\ref{parametrization2}). In the overlapping regions,
\begin{equation}
\chi_\alpha^{NN} = e^{i\xi_1} \chi_\alpha^{SN},\chi_\alpha^{NS} = e^{i(\xi_1-\xi_2)} \chi_\alpha^{SN} \,\,\, \textrm{etc.}
\end{equation}
A complex scalar field on $S^2_N\times S^2_N$ is a function of $\chi_\alpha^{NN}$
\begin{eqnarray}
\Phi_{NN} &=& \sum c_{n_\alpha n_\alpha'}\left[\left(\bar{\chi}_1^{NN}\right)^{n'_1}\left(\bar{\chi}_2^{NN}\right)^{n'_2}\left(\bar{\chi}_3^{NN}\right)^{n'_3}
\left(\bar{\chi}_4^{NN}\right)^{n'_4}\right. \nonumber\\
 && \left.\left(\chi_1^{NN}\right)^{n_1}\left(\chi_2^{NN}\right)^{n_2}\left(\chi_3^{NN}\right)^{n_3}\left(\chi_4^{NN}\right)^{n_4}\right]
\end{eqnarray}
where $S^2_N \equiv S^2-\{S\}$. On another patch,  say,  $S^2_S\times S^2_S$ $(S^2_S\equiv S^2-\{N\})$, the field is a function of $\chi_\alpha^{SS}$
\begin{eqnarray}
\Phi_{SS} &=& \sum c_{n_\alpha n_\alpha'}\left[\left(\bar{\chi}_1^{SS}\right)^{n'_1}\left(\bar{\chi}_2^{SS}\right)^{n'_2}\left(\bar{\chi}_3^{SS}\right)^{n'_3}
\left(\bar{\chi}_4^{SS}\right)^{n'_4}\right. \nonumber \\ &&\left.\left(\chi_1^{SS}\right)^{n_1}\left(\chi_2^{SS}\right)^{n_2}\left(\chi_3^{SS}\right)^{n_3}\left(\chi_4^{SS}\right)^{n_4}\right].
\end{eqnarray}
If $n'_1+n'_2+n'_3+n'_4-n_1-n_2-n_3-n_4= \kappa $= fixed, then in the overlapping region $\left(S^2_N\times S^2_N \right)\cap\left( S^2_S\times S^2_S\right)$, $\Phi_{NN}$ is related to $\phi_{SS}$ as
\begin{equation}
\Phi_{NN} = e^{i\kappa(\xi_1+\xi_2)} \Phi_{SS}.
\end{equation}
The phase $ e^{i\kappa(\xi_1+\xi_2)}$  in the above equation can be identified as the gauge transformation relating $\Phi_{NN}$ to $\Phi_{SS}$. This gauge transformation arises from a gauge field $A_\mu$ with 
\begin{eqnarray}
&\quad\quad  A_\mu^{NN} = i\kappa \bar{\chi}^{NN}_\alpha\left(\partial_\mu \chi_\alpha^{NN}\right) \quad \mathrm{on} \,\,\, S^2_N\times S^2_N, \nonumber \\
&A_\mu^{SS} = i\kappa\bar{\chi}^{SS}_\alpha\left(\partial_\mu \chi_\alpha^{SS}\right) \quad \mathrm{on} \,\,\, S^2_S\times S^2_S , \\
&\mathrm{and}\,\,\, \,\,\,\, A^{NN}_\mu = A^{SS}_\mu - i \mathcal{G}\left(\partial_\mu \mathcal{G}^{-1}\right)\quad \mathrm{on}\,\,\, S^2_N \times  S^2_N \cap S^2_S\times  S^2_S  \nonumber
\end{eqnarray}
where $\mathcal{G}=e^{i\kappa(\xi_1+\xi_2)}$ and $\mu=\theta_1,\xi_1,\theta_2,\xi_2$. We can explicitly compute $A_\mu$:
\begin{eqnarray}
&&\hspace{-1cm}A_{\theta_1}^{NN}=0, \quad A_{\xi_1}^{NN}=\frac{\kappa}{2} (1-\cos \theta_1),\quad A_{\theta_2}^{NN}=0,\quad A_{\xi_2}^{NN}=\frac{\kappa}{2} (1-\cos \theta_2), \\
&&\hspace{-1cm}A_{\theta_1}^{SS}=0, \quad A_{\xi_1}^{SS}=-\frac{\kappa}{2} (1+\cos \theta_1),\quad A_{\theta_2}^{SS}=0,\quad A_{\xi_2}^{SS}=-\frac{\kappa}{2} (1+\cos \theta_2).
\end{eqnarray}
The connection one-forms are
\begin{eqnarray}
A_{NN}&=& A^{NN}_{\theta_1} d\theta_1 + A^{NN}_{\xi_1} d\xi_1+A^{NN}_{\theta_2} d\theta_2 + A^{NN}_{\xi_2} d\xi_2\nonumber\\
& =&\frac{\kappa}{2}(1-\cos \theta_1) d\xi_1+\frac{\kappa}{2}(1-\cos \theta_2) d\xi_2, \label{form_north1}\\ \nonumber \\
 A_{SS}&=& A^{SS}_{\theta_1} d\theta_1 + A^{SS}_{\xi_1} d\xi_1+A^{SS}_{\theta_2} d\theta_2 + A^{SS}_{\xi_2} d\xi_2\nonumber\\
& =& -\frac{\kappa}{2}(1+\cos \theta_1) d\xi_1-\frac{\kappa}{2}(1+\cos \theta_2) d\xi_2.\label{form_south1}
\end{eqnarray}
In the overlapping region $\left(S^2_N\times S^2_N \right)\cap\left( S^2_S\times S^2_S\right)$, they  are related as 
\begin{eqnarray}
 A_{NN}-A_{SS}= \kappa (d\xi_1+d\xi_2)=\kappa d(\xi_1+\xi_2)
\end{eqnarray}
where $\kappa$ is  integer.

By similar arguments, we can show that
\begin{eqnarray}
A_{NS}&=&\frac{\kappa}{2}(1-\cos \theta_1) d\xi_1-\frac{\kappa}{2}(1+\cos \theta_2) d\xi_2, \\ \nonumber \\
 A_{SN}&=& -\frac{\kappa}{2}(1+\cos \theta_1) d\xi_1+\frac{\kappa}{2}(1-\cos \theta_2) d\xi_2
\end{eqnarray}
and in the overlapping regions they are related as
\begin{equation}
 A_{NN}-A_{NS}= \kappa d\xi_2,\quad  A_{NS}-A_{SN}= \kappa d(\xi_1-\xi_2) \quad \textrm{etc.}
\end{equation}

We denote the space of the complex scalar fields (with $\kappa=$ fixed) as $\mathcal{H}_\kappa$. Any element $\Phi$ of $\mathcal{H}_\kappa$ is an eigenfunction of the operator $K_0$:
\begin{eqnarray}
K_0=\frac{1}{2}\sum_{\alpha=1}^4 \left(\bar{z}_{\alpha}\frac{\partial}{\partial \bar{z}_{\alpha}} - z_{\alpha}\frac{\partial}{\partial z_{\alpha}}\right), \quad K_0\Phi=\frac{\kappa}{2} \Phi, \quad \kappa=\mathrm{fixed}.
\label{topo_charge_1}
\end{eqnarray}
The differential operators $W_i$ and $Y_i$ defined as 
\begin{eqnarray}\left.
\begin{array}{llll}
&&W_i=-i\epsilon_{ijk} \omega_j \frac{\partial}{\partial \omega_k }, \quad Y_i=-i\epsilon_{ijk} y_j \frac{\partial}{\partial y_k }, \\ \\
&&W^2 =W_iW_i,  \quad Y^2 = Y_iY_i, \quad W^2=Y^2
\end{array}\right.
\end{eqnarray}
satisfy
\begin{eqnarray}\left.\begin{array}{cccc}
&[W_i, W_j] =i\epsilon_{ijk} W_k, \quad [Y_i, Y_j]=i\epsilon_{ijk}Y_k, \quad [W_i, Y_j]=0,\quad [W_i,K_0]=0,\\ \\
&[W^2,W_i]=0=[W^2,Y_i],\quad [Y_i,K_0]=0,\quad [W^2,K_0]=0.
\end{array}\right.
\end{eqnarray}
These  differential operators  map $\mathcal{H}_\kappa \rightarrow \mathcal{H}_\kappa$ and are the generators of rotations in $\mathcal{H}_\kappa$.  We can choose the simultaneous eigenfunctions of $W_3, Y_3$ and $W^2$ as  basis vectors of  $\mathcal{H}_\kappa$. These are also the eigenfunctions of $K_0$  with fixed eigenvalue $\kappa$. So $\mathcal{H}_\kappa$ is spanned by functions (basis vectors) $\phi^{j, m^\prime}_{\kappa, m}$
\begin{eqnarray}\left.
\begin{array}{lll}
&&W^2 \Phi^{j, m^\prime}_{\kappa, m}= j(j+1) \Phi^{j, m^\prime}_{\kappa, m}=Y^2 \Phi^{j, m^\prime}_{\kappa, m},\\ \\
&&W_3 \Phi^{j, m^\prime}_{\kappa, m}= m \Phi^{j, m^\prime}_{\kappa, m},    \quad   \quad \quad\quad \quad\quad\quad \quad\quad\quad -j\leq m\leq j,\\ \\
&& Y_3 \Phi^{j, m^\prime}_{\kappa, m}= m^\prime \Phi^{j, m^\prime}_{\kappa, m},    \quad \quad \quad \quad\quad \quad \quad \quad\quad\quad -j\leq m^\prime\leq j,\\ \\
&&K_0\Phi^{j, m^\prime}_{\kappa, m}=\frac{ \kappa}{2} \Phi^{j, m^\prime}_{\kappa, m},\quad   \quad \quad\quad \quad \quad \quad \quad \quad\quad \quad \kappa=\mathrm{fixed}. 
\end{array}\right.
\end{eqnarray}
Below we construct the basis functions explicitly, which will give us the allowed values of $j$.

The function $f=N_{\tilde{l}\tilde{n}}\bar{z}_1^{\tilde{l}}z_4^{\tilde{n}}$ with $(\tilde{l}-\tilde{n})=\kappa=$ fixed,   satisfies the following relations
\begin{equation} \left.
\begin{array}{lll}
&&W_+ f \equiv (W_1+ iW_2) f= 0, \\ \\
&& W_3 f= \frac{\tilde{l}+\tilde{n}}{2} f , \\ \\
&&Y_+ f \equiv (Y_1+ iY_2) f= 0, \\ \\
&& Y_3 f= \frac{\tilde{l}+\tilde{n}}{2} f , \\ \\
 &&K_0 f= \frac{\tilde{l} - \tilde{n}}{2} f=\frac{\kappa}{2} f, 
\end{array}\right\}\quad \tilde{n},\tilde{l}\in \mathbb{N}_0,\quad \kappa=\mathrm{fixed}.
\end{equation}
So $f$ is the highest weight vector of the two $SU(2)$'s generated by $W_i$ and $Y_i$ in the irreducible representation with $j=\frac{\tilde{l}+\tilde{n}}{2}$.
From this starting point,  we can generate the other basis vectors which span $\mathcal{H}_\kappa$. Denoting  this  $f$ as $\Phi^{j, j}_{\kappa,j}$, the other basis vectors can be obtained by repeated application of $W_- (\equiv W_1-iW_2)$ and $Y_- (\equiv Y_1-iY_2)$ on $\Phi^{j, j}_{\kappa,j}$:
\begin{eqnarray}
&&(W_-)^{(j-m)}\Phi^{j, j}_{\kappa, j}  =q^{j,j}_{\kappa m} \Phi^{j,j}_{\kappa,m}, \\ \nonumber \\
&&(Y_-)^{(j-m^{\prime})} \Phi^{j, j}_{\kappa, j} =q^{j,m^\prime}_{\kappa j} \Phi^{j,m^\prime}_{\kappa j}, \\ \nonumber \\
&&(W_-)^{(j-m)}(Y_-)^{(j-m^{\prime})} \Phi^{j, j}_{\kappa, j} = q^{j,m^\prime}_{\kappa m} \Phi^{j,m^\prime}_{\kappa m}.
\end{eqnarray}
In the above, $ q^{j,m^\prime}_{\kappa m}$ are constants.

As $\tilde{n}$ and $\tilde{l}$ are arbitrary positive integers satisfying  $\tilde{n}-\tilde{l}=\kappa$,  the allowed values of $j$ are 
\begin{equation}
j=\frac{\tilde{n}+\tilde{l}}{2}, \quad j=\frac{\kappa}{2}, \frac{\kappa}{2}+1, \frac{\kappa}{2}+2......\infty.
\end{equation}
So $\mathcal{H}_\kappa$ is spanned by the basis
\begin{equation}
\left\{\Phi^{j, m^\prime}_{\kappa, m}\right\} \quad\quad\mathrm{with}\quad -j \leq m \leq j, \quad -j \leq m^\prime \leq j, \quad j=\frac{\kappa}{2}, \frac{\kappa}{2}+1, \frac{\kappa}{2}+2.....\infty
\end{equation} 
and an arbitrary element $\Phi$ of $\mathcal{H}_\kappa$ can be expressed as
\begin{equation}
\Phi = \sum_{j=\frac{\kappa}{2}}^\infty\sum_{m=-j}^j c^{j, m^\prime}_{\kappa, m}\Phi^{j, m^\prime}_{\kappa, m}, \quad\quad   c^{j, m^\prime}_{\kappa, m} \in \mathbb{C}.
\end{equation}
These $\Phi$'s are thus  the sections of the $U(1)$ fibre bundle with the {\it topological charge} $\kappa$.

\section{Monopoles from K-K Reduction $ X^5 \rightarrow S^2 \times S^2$}{\label{kaluza_klein_sec}}

Interestingly, the principal fibre bundle of the previous discussion can be obtained by the Kaluza-Klein reduction of  the metric on  $X^5$. 
The metric (\ref{metric3}) can be written as $ds_{X^5}^2= \tilde{g}_{ab} d\eta^a d\eta^b$ where
\begin{equation}\tilde{g}_{ab}=\left(
\begin{array}{ccccc}
\frac{1}{4}&0&0&0&0 \\
0&\frac{1}{4}\sin^2 \theta_1+\frac{2}{3}\sin^4\frac{\theta_1}{2}&0&\frac{2}{3}\sin^2\frac{\theta_1}{2}\sin^2\frac{\theta_2}{2}&\frac{2}{3}\sin^2\frac{\theta_1}{2}\\
0&0&\frac{1}{4}&0&0\\
0&\frac{2}{3}\sin^2\frac{\theta_1}{2}\sin^2\frac{\theta_2}{2}&0&\frac{1}{4}\sin^2 \theta_2+\frac{2}{3}\sin^4\frac{\theta_2}{2}&\frac{2}{3}\sin^2\frac{\theta_2}{2}\\
0&\frac{2}{3}\sin^2\frac{\theta_1}{2}&0&\frac{2}{3}\sin^2\frac{\theta_2}{2}&\frac{2}{3}
\end{array}\right);\quad 
\label{metric12}
\end{equation}
where $\eta^1=\theta_1, \eta^2=\xi_1, \eta_3=\theta_2,\eta_4=\xi_2, \eta^5=\psi$.
This metric satisfies
\begin{eqnarray}
\tilde{R}_{ab}=\frac{8}{3}\tilde{g}_{ab}
\end{eqnarray}
and is a solution of the Einstein equations with a positive cosmological constant.

The metric (\ref{metric12}) already has the convenient   Kaluza-Klein form
\begin{equation}\tilde{g}_{ab}=\left(
\begin{array}{cccc}
g_{\mu\nu}+g A_\mu A_\nu &g A_\mu\\
g A_\nu & g
\end{array}\right), \quad 
g_{\mu\nu}=\left(
\begin{array}{cccc}
\frac{1}{4}&0&0&0\\
0& \frac{1}{4}\sin^2 \theta_1&0&0\\
0&0&\frac{1}{4}&0\\
0&0&0&\frac{1}{4}\sin^2 \theta_2
\end{array}\right),
\label{kk_metric}
\end{equation}
with one extra compact dimension $\psi$ and the dilaton $g$ set to $\frac{2}{3}$.
$g_{\mu\nu}$ is  the metric on $S^2_N \times S^2_N$ and satisfies 
\begin{eqnarray}
R_{\mu\nu}=4g_{\mu\nu}.
\end{eqnarray}

Inspecting (\ref{metric12}) immediately tells us that the gauge fields on this local patch are 
\begin{eqnarray}
A^{NN}_{\theta_1}=0, \quad\quad  A^{NN}_{\xi_1}= \sin^2 \frac{\theta_1}{2}=\frac{1}{2}(1-\cos \theta_1),\nonumber \\
A^{NN}_{\theta_2}=0, \quad \quad A^{NN}_{\xi_2}= \sin^2 \frac{\theta_2}{2}=\frac{1}{2}(1-\cos \theta_2).
\label{gaugefields2}
\end{eqnarray}

Similarly we get on $S^2_S \times S^2_S$ 
\begin{eqnarray}
&& A^{SS}_{\theta_1}=0,\quad  A^{SS}_{\xi_1}=-\frac{1}{2}(1+\cos \theta_1), \nonumber \\
&& A^{SS}_{\theta_2}=0,\quad A^{SS}_{\xi_2}= -\frac{1}{2}(1+\cos \theta_2).
\end{eqnarray}
$A^{NN}_\mu$ and $A^{SS}_\mu$ are related by the gauge transformation 
\begin{equation}
A^{NN}_{\xi_1}=A^{SS}_{\xi_1} -i \mathcal{G} \partial_{\xi_1} \mathcal{G}^{-1}, \quad A^{NN}_{\xi_2}=A^{SS}_{\xi_2} -i\mathcal{G}\partial_{\xi_2}\mathcal{G}^{-1}
\end{equation}
where $\mathcal{G}=e^{i(\xi_1+\xi_2)} \in U(1)$.
This is a version of the Kaluza-Klein monopole \cite{{ Sorkin:1983ns},{Gross:1983hb},{RandjbarDaemi:1982hi}}.

Notice that the metric on $X^5$ is in the K-K form. Further  both $X^5$ and $S^2 \times S^2$ are Einstein manifolds.  In general if there is a Kaluza-Klein reduction of $\mathcal{M}^{d+1}$ to $\mathcal{M}^{d}$ where both $\mathcal{M}^{d+1}$ and $\mathcal{M}^{d}$ are compact Einstein manifolds, then
as shown in \cite{Acharyya:2013hga},  the ``Einstein condition''  leads to stringent conditions on the gauge field action.  Let $\tilde{g}_{ab}$ be the metric on $\mathcal{M}^{d+1}$, $g_{\mu\nu}$ the metric on $\mathcal{M}^{d}$ and the dilaton $\tilde{g}_{d+1,d+1}( \equiv g$) a constant.
%
%
%
%
%
%
The Einstein condition 
\begin{equation}
\tilde{R}_{ab}= c\tilde{g}_{ab}, \quad R_{\mu\nu}= c_0 g_{\mu\nu}
\end{equation}
forces the gauge fields to satisfy
\begin{eqnarray} \label{cond3}
&&F^{\sigma\beta}F_{\sigma\beta} =\frac{2}{g}(c_0 - c)d,\\
\textrm{with} && c_0= \frac{d+2}{d}c. \label{reln1}
\end{eqnarray}
Moreover as $\mathcal{M}^{d}$ is compact, (\ref{cond3}) implies that  the electromagnetic action 
\begin{equation}
S_{EM} = \int d(Vol) \left(\frac{1}{4}F^{\sigma\beta}F_{\sigma\beta}\right)=\frac{c}{g} Volume
\end{equation}
is finite.

 From  the metric (\ref{metric12}) on $X^5$ we get 
\begin{equation}
d=4, \quad c=\frac{8}{3}, \quad c_0= 4, \quad g= \frac{2}{3}.
\end{equation}
So (\ref{reln1}) is satisfied.

$F_{\mu\nu}$ can be computed explicitly from (\ref{gaugefields2}):
\begin{eqnarray}
 F_{\mu\nu}&=&\left(\begin{array}{ccccc}
0&\frac{\sin\theta_1}{2}&0&0\\
-\frac{\sin\theta_1}{2}&0&0&0\\
0&0&0&\frac{\sin\theta_2}{2}\\
0&0&-\frac{\sin\theta_2}{2}&0
\end{array}\right).
\end{eqnarray}
This leads to
\begin{eqnarray}
F^{\mu\nu}F_{\mu\nu} &=& -F^{\mu\nu}F_{\nu\mu}=16
\end{eqnarray}
 which is in accordance to (\ref{cond3})
and the electromagnetic action is 
\begin{equation}
S_{EM} =\frac{c}{g} Volume_{S^2 \times S^2}=4\pi^2.
\end{equation}

\section{Fuzzy Fibre Bundle}\label{fuzzy_section}

 In the spirit of \cite{Grosse:1995jt, Acharyya:2013hga}, we want to describe the fuzzy conifold $Y_F^6$ and its monopole bundles. Here we will not attempt to provide a mathematically precise definition of a fuzzy space, as  universally acceptable definition of a fuzzy space does not seem to exist in the known literature.  We are interested in only a specific class of fuzzy spaces and to deal with such  spaces, we will state ``working definitions" wherever necessary.

Many fuzzy spaces  have a description in terms of  bosonic oscillators $\hat{a}_\alpha,\hat{a}_\alpha^\dagger$ (or more precisely, specific operator functions of $\hat{a}_\alpha,\hat{a}_\alpha^\dagger$). For example, the quantized  $n$--dimensional complex plane $\mathbb{C}^n_F$ is represented by $n$ independent  bosonic oscillators
\begin{equation}
[\hat{a}_\alpha, \hat{a}_\beta^\dagger]=\delta_{\alpha\beta},\quad  \,\,\, \, \alpha,\beta=1,2,3\ldots n,
\end{equation}
which  act on the Hilbert space of the $n$ bosonic oscillators. Restricting   to appropriate subspaces of this Hilbert space gives us other fuzzy spaces like  $S^{2n-1}_F$, $\mathbb{CP}_F^n$ etc. \cite{Balachandran:2005ew,Balachandran:2001dd, Dolan:2006tx}. On these spaces the discrete analogue of the classical objects like  instantons, solitons, monopoles and the like are of interest (for example see \cite{ {Hoppe:1982},{Madore:1991bw},{Grosse:1992bm}, {Baez:1998he},{Balachandran:1999hx},{Vaidya:2001rf},{Nekrasov:1998ss},{Acharyya:2011bx}}). 

 The spaces of our interest are the fuzzy conifold $Y_F^6$ and  fuzzy spheres.
 A ``Hopf-like" construction  relates the fuzzy conifold $Y^6_F$ with $S_F^2 \times S_F^2$. This is the fuzzy analogue of the map (\ref{map1}). On these fuzzy spaces, the standard differential geometric tools are unavailable. Nevertheless some topological  information can be extracted indirectly by studying the  complex line bundles, in particular by using the group action of $SU(2)$ \cite{Grosse:1995jt,Acharyya:2013hga}. 
We will adapt  this technique to describe the fuzzy fibration $X^5_F \rightarrow S_F^2\times S_F^2$, construct the corresponding line bundles and identify the monopole charges.

\subsection{The Fuzzy Conifold $Y_F^6$ and the Base $X_F^5$ }
$\mathbb{C}^4_F$  is described by the algebra of  four independent oscillators
\begin{equation}
\left[\hat{a}_\alpha, \hat{a}_\beta^\dagger\right]=\delta_{\alpha\beta}, \quad\quad \mathrm{where} \quad \alpha,\beta=1,2,3,4,
\end{equation}
which act on a space $\mathcal{F}$  spanned by the eigenstates of the {\it number} operators $\hat{N}_\alpha \equiv \hat{a}_\alpha^\dagger \hat{a}_\alpha$:
\begin{equation}
\mathcal{F}= Span\left\{|n_1,n_2, n_3, n_4 \rangle: n_\alpha=0,1,2\ldots\right\}.
\end{equation}
The total number operator is $$\hat{N} \equiv\sum_{\alpha=1}^4\hat{N}_\alpha.$$

In analogy with  (\ref{conifold_1})
let us define the operator $\hat{\mathcal{O}}_4$ as
\begin{equation}
\hat{\mathcal{O}}_4 \equiv \hat{a}_1 \hat{a}_4-\hat{a}_2\hat{a}_3
\end{equation}
which has as its kernel 
\begin{equation}
\ker(\hat{\mathcal{O}}_4) = \left\{|n_1,n_2, n_3,n_4 \rangle: \,\, \hat{\mathcal{O}}|n_1,n_2, n_3,n_4 \rangle=0 \right\}.
\end{equation}

The usual conifold (\ref{conifold_1}) is defined by the commutative algebra of polynomial functions of $z_\alpha$, subject to the condition 
$z_1 z_4-z_2 z_3=0$. To get the noncommutative analogue of the conifold let us recall the argument we gave in \cite{Acharyya:2013hga}.
The algebra of the polynomial functions of $\hat{a}_\alpha$, subject to the condition $ \hat{a}_1\hat{a}_4-\hat{a}_2\hat{a}_3=0$ as it stands is a commutative algebra. 
However the $\hat{a}_\alpha$'s  are infinite dimensional operators and under the $\ast$--operation (where $(\hat{a}_\alpha)^\ast \equiv \hat{a}_\alpha^\dagger$) this algebra as a star algebra is noncommutative. 
Therefore we interpret the condition  $\hat{a}_1\hat{a}_4-\hat{a}_2\hat{a}_3=0$ as a constraint on the set of admissible states in the bosonic Hilbert space of the $\hat{a}_\alpha$'s.

Thus our operational  definition of the fuzzy  conifold $Y_F^6$ is the restriction of the action of the operators $\hat{a}_{\alpha}$  (and polynomial operator functions of $\hat{a}_\alpha$) to $\ker(\hat{\mathcal{O}}_4)$. Defining a fuzzy space by such a restriction is not new (for example see \cite{Balachandran:2005ew, Acharyya:2013hga}).

It is easy to see why this definition of $Y_F^6$ is appropriate. Let $|z_1,z_2,z_3,z_4 \rangle$ be the  standard coherent states of 4--dimensional oscillator
\begin{equation}
|z_1,z_2,z_3,z_4 \rangle=N(z_\alpha, \bar{z}_\alpha) e^{z_\alpha \hat{a}_\alpha^\dagger}|0,0,0,0\rangle.
\label{coherent_states2}
\end{equation}
Then 
\begin{equation}
\hat{\mathcal{O}}_4 |z_1,z_2,z_3,z_4 \rangle = \left(z_1 z_4-z_2 z_3\right)|z_1,z_2,z_3 \rangle.
\end{equation}
This implies that if the coherent state $|z_1,z_2,z_3,z_4 \rangle \in \ker (\hat{\mathcal{O}}_4)$, the complex numbers $z_\alpha$ satisfy (\ref{conifold_1}).

Let us define
\begin{equation}
\hat{\chi}_\alpha = \hat{a}_\alpha \frac{1}{\sqrt{\hat{N}}}, \quad \hat{\chi}_\alpha^\dagger =  \frac{1}{\sqrt{\hat{N}}}\hat{a}_\alpha^\dagger
\end{equation}
which satisfy 
\begin{equation}
\hat{\chi}_\alpha^\dagger \hat{\chi}_\alpha = 1.
\label{seven_sphere}
\end{equation}
$\hat{\chi}_\alpha$ is well-defined if we exclude the state $|0,0,0,0 \rangle$  from the domain of definition of $\hat{\chi}_\alpha$.
Then (\ref{seven_sphere}) gives us the fuzzy 7--sphere $S_F^7$. 
The operator
\begin{equation}
\hat{\mathcal{O}}_4' \equiv\hat{ \chi}_1\hat{ \chi}_4-\hat{ \chi}_2\hat{ \chi}_3 = \frac{1}{\sqrt{(\hat{N}+1)(\hat{N}+2)}} \hat{\mathcal{O}}_4
\end{equation}
obviously vanishes on $\ker(\hat{\mathcal{O}}_4)$. The restriction of  $\hat{\chi}_\alpha$ to $\ker(\hat{\mathcal{O}}_4)$  defines  $X^5_F$. 
This is the fuzzy version of $X^5$ and can be informally thought of as  the intersection of $Y^6_F$ with $S_F^7$.

The continuum $X^5$ can be recovered in the limit $\hat{N} \rightarrow \infty$ of $X_F^5$. It can be easily seen  using the coherent states (\ref{coherent_states2}).

\subsection{$S_F^2\times S^2_F$ and the Noncommutative Fibre Bundle}{\label{fuzzy_fibre_bundle}}

Using the matrices (\ref{matrices}),  the fuzzy analogue of the map (\ref{map1}) is
\begin{equation}\left. \begin{array}{lll}
\hat{\omega}_i = \hat{\chi}^\dagger A_i \hat{\chi} = \frac{1}{\hat{N}}\hat{L}_i, &&
\hat{L}_i=\hat{a}^\dagger_\alpha (A_i)_{\alpha\beta} \hat{a}_\beta, \\ \\
 \hat{y}_i= \hat{\chi}^\dagger B_i \hat{\chi} = \frac{1}{\hat{N}}\hat{S}_i,&& \hat{S}_i=\hat{a}^\dagger_\alpha (B_i)_{\alpha\beta} \hat{a}_\beta,
\end{array}\right.
\quad \quad\quad \quad \textrm{where} \quad\hat{\chi}=\left(\begin{array}{lll}
\hat{\chi}_1 \\
\hat{\chi}_2\\
\hat{\chi}_3 \\
\hat{\chi}_4
\end{array}\right).
\label{map}
\end{equation}
This preserves  the Lie algebra of the matrices $A_i$ and $B_i$:
\begin{eqnarray}
&&\left[\hat{L}_i, \hat{L}_j\right]=i\epsilon_{ijk}\hat{L}_k, \quad \left[\hat{S}_i,\hat{S}_j\right]=i\epsilon_{ijk}\hat{S}_k,\nonumber \\
&& \left[\hat{L}_i,\hat{S}_j\right]=,\quad \left[\hat{L}_i, \hat{N}\right]= 0 =\left[\hat{S}_i, \hat{N}\right].
\end{eqnarray}
The above algebraic structure is that of $SU(2) \times SU(2)$ and the Casimirs are 
\begin{eqnarray}
 \hat{L}_i\hat{L}_i = \frac{N}{2}\left(\frac{N}{2}+1\right) - \hat{\mathcal{O}_4}^\dagger \hat{\mathcal{O}_4},\nonumber \\
 \hat{S}_i\hat{S}_i= \frac{N}{2}\left(\frac{N}{2}+1\right) - \hat{\mathcal{O}_4}^\dagger \hat{\mathcal{O}_4}.
\end{eqnarray}
Those representations of $SU(2)\times SU(2)$ which have the same value for the Casimirs are sometimes denoted by $SU(2)\star SU(2)$ \cite{louck},
 with the Casimirs denoted by a single operator 
\begin{equation}
\hat{C} \equiv \hat{L}_i\hat{L}_i = \hat{S}_i\hat{S}_i.
\end{equation}

As $\hat{L}_i$ and $\hat{S}_i$ commute with $\hat{N}$, it is easy to see that
\begin{eqnarray}
&&\hat{\omega}_i\hat{\omega}_i = \frac{1}{\hat{N}^2 }\hat{L}_i\hat{L}_i =  \frac{1}{2}\left(\frac{1}{2}+\frac{1}{\hat{N}}\right) - \frac{1}{\hat{N}^2}\hat{\mathcal{O}_4}^\dagger \hat{\mathcal{O}_4}, \\
&&\hat{y}_i\hat{y}_i = \frac{1}{\hat{N}^2 }\hat{S}_i\hat{S}_i ==  \frac{1}{2}\left(\frac{1}{2}+\frac{1}{\hat{N}}\right) - \frac{1}{\hat{N}^2}\hat{\mathcal{O}_4}^\dagger \hat{\mathcal{O}_4}. 
\end{eqnarray}

In $\ker(\hat{\mathcal{O}}_4)$, 
\begin{eqnarray}
\hat{C}|_{\ker(\hat{\mathcal{O}}_4)} = \frac{N}{2}\left(\frac{N}{2}+1\right), \quad \hat{\omega}_i\hat{\omega}_i|_{\ker(\hat{\mathcal{O}}_4)}  = \hat{y}_i\hat{y}_i|_{\ker(\hat{\mathcal{O}}_4)} =  \frac{1}{2}\left(\frac{1}{2}+\frac{1}{\hat{N}}\right).
\label{casimir_fuzzy678}
\end{eqnarray}
This is the fuzzy analogue of (\ref{casimir1}) and (\ref{casimir678}).

It is convenient to decompose $\mathcal{F}$ into subspaces $\mathcal{F}_n$ in which $\hat{N} $ takes a fixed value $n$:
\begin{equation}
\mathcal{F}_n= \left\{|n_1,n_2, n_3,n_4 \rangle: n_1+n_2+n_3+n_4=n \right\}, \quad \mathcal{F}=\oplus_n \mathcal{F}_n.
\end{equation}
 The dimension of $\mathcal{F}_n$ (number of states) is 
\begin{equation}
d_n= \frac{(n+1)(n+2)(n+3)}{6}.
\end{equation}

Let $\tilde{\mathcal{F}}_n$  be the subspace of $\mathcal{F}_n$ defined as
\begin{equation}
\tilde{\mathcal{F}}_n= \mathcal{F}_n \cap \ker(\hat{\mathcal{O}}_4).
\end{equation}
In $\tilde{\mathcal{F}}_n$,
\begin{equation}
\hat{\omega}_i\hat{\omega}_i=\hat{y}_i\hat{y}_i = \left(1+\frac{1}{n}\right)\mathbb{1}.
\end{equation}
Thus the algebra generated by the $\hat{\omega}_i$'s and  $\hat{y}_i$'s restricted to $\tilde{\mathcal{F}}_n$ is that of $S^2_F \times S^2_F$ and  (\ref{map}) is the fuzzy map $X^5_F \rightarrow S^2_F \times S^2_F$. 
In the commutative limit  $n \rightarrow \infty$, we recover $S^2\times S^2$ with each sphere having radius $\frac{1}{2}$.

When restricted to $\tilde{\mathcal{F}}_n$,  $\hat{C}$ takes the fixed value $\frac{n}{2}\left(\frac{n}{2}+1\right)$ and $\tilde{\mathcal{F}}_n$ is the carrier space for a finite dimensional UIR of the $SU(2) \star SU(2)$.

The states of $SU(2) \star SU(2)$ are labeled as $|\frac{n}{2},\tilde{j}, m_1, m_2 \rangle$ where 
\begin{eqnarray}
&&\hat{C} |\frac{n}{2}, \tilde{j}, m_1, m_2 \rangle = \tilde{j}(\tilde{j}+1) |\frac{n}{2},\tilde{j}, m_1, m_2 \rangle , \nonumber \\
 && N |\frac{n}{2}, \tilde{j}, m_1, m_2 \rangle = n |\frac{n}{2}, \tilde{j}, m_1, m_2 \rangle, \nonumber\\
&& \hat{\omega}_3 |\frac{n}{2}, \tilde{j}, m_1, m_2 \rangle = m_1 |\frac{n}{2}, \tilde{j}, m_1, m_2 \rangle,\nonumber\\
&& \hat{y}_3 |\frac{n}{2},\tilde{j}, m_1, m_2 \rangle =m_2 |\frac{n}{2},\tilde{j}, m_1, m_2 \rangle.
\label{states_su2_su2}
\end{eqnarray}

Comparing (\ref{casimir_fuzzy678}) and (\ref{states_su2_su2}) we get $\tilde{j}= \frac{n}{2}$. 
For a given value of $\tilde{j}$ (i.e. for a fixed value of $n$), $m_1$ and $m_2$ take values 
\begin{equation}
m_1=-\tilde{j}, -\tilde{j}+1,.....\tilde{j}-1,\tilde{j}, \quad m_2=-\tilde{j}, -\tilde{j}+1,.....\tilde{j}-1, \tilde{j}.
\end{equation}
So there are $(2\tilde{j}+1) \times (2\tilde{j}+1)$ states and hence the dimension $p_n$ of  $\tilde{\mathcal{F}} _n$ is
\begin{equation}
 p_n=\left(2\tilde{j}+1\right)^2 = (n+1)^2.
\end{equation}

We will exploit the group theoretic properties of $SU(2) \star SU(2)$ to characterize the fuzzy fibre bundle. The strategy here is the same as in  \cite{Grosse:1995jt, Acharyya:2013hga}. 
Let $\mathcal{H}_{nl}$ be the space of linear operators which map  $\tilde{\mathcal{F}}_n$ to $ \tilde{\mathcal{F}}_l$:
\begin{equation}
\hat{\Phi}: \tilde{\mathcal{F}}_n \rightarrow \tilde{\mathcal{F}}_l, \quad \hat{\Phi} \in \mathcal{H}_{nl}.
\end{equation} 
These operators $\hat{\Phi}$ are represented by $(l+1)^2 \times (n+1)^2$ rectangular matrices.

In particular, $\mathcal{H}_{nn}$ are $\left[(n+1)^2\right]^2$ dimensional noncommutative algebras $\mathcal{A}_n$. 
In $\mathcal{H}_{nn}$, the two $SU(2)$ rotations are generated by the adjoint action of $L_i^{(n)}$ and $S_i^{(n)}$ respectively:
\begin{equation}\left.\begin{array}{lll}
Ad(\hat{L}_i)\hat{\Phi}\equiv\hat{W}_i\hat{\Phi}\equiv[\hat{L}^{(n)}_i, \hat{\Phi}],\\
Ad(\hat{S}_i)\hat{\Phi}\equiv\hat{Y}_i\hat{\Phi}\equiv[\hat{S}^{(n)}_i, \hat{\Phi}],
\end{array}\right. \quad\quad \hat{\Phi} \in \mathcal{H}_{nn}.
\end{equation}
The operators $\hat{W}_i$ and $\hat{Y}_i$ satisfy 
\begin{eqnarray} \left. \begin{array}{lll}
&\left[\hat{W}_i, \hat{W}_j\right]= i\epsilon_{ijk} \hat{W}_k,\quad \left[\hat{Y}_i, \hat{Y}_j\right]=i\epsilon_{ijk} \hat{Y}_k, \quad \left[\hat{W}_i, \hat{Y}_j\right]=0, \\ \\
&\hat{W}^2 \equiv \hat{W}_i \hat{W}_i, \quad \hat{Y}^2 \equiv \hat{Y}_i \hat{Y}_i, \quad  \hat{W}^2 =\hat{Y}^2,\quad  \left[\hat{W}^2,\hat{W}_i\right] =0=\left[\hat{Y}^2,\hat{Y}_i\right].
\end{array}  \quad\right.
\end{eqnarray}
$\hat{L}_i^{(n)}$ and $\hat{S}_i^{(n)}$ are the restriction of the operators $\hat{L}_i$ and $\hat{S}_i$ in the subspace $\tilde{\mathcal{F}}_n$. 
As $\tilde{\mathcal{F}}_n$ is the carrier space of $(n+1)^2$-dimensional UIR of the $SU(2)\star SU(2)$, the matrices $\hat{L}_i^{(n)}$ and $\hat{S}_i^{(n)}$  are $(n+1)^2 \times (n+1)^2$ matrices.

The space $\mathcal{H}_{nl}$ is a noncommutative bimodule: it is a left $\mathcal{A}_l$--module and a right $\mathcal{A}_n$--module. 
In the bimodule $ \mathcal{H}_{nl}$, $\hat{W}_i$ and $\hat{Y}_i$ act as 
\begin{eqnarray} \left.\begin{array}{lll}
&&\hat{W}_i\hat{\Phi} = \hat{L}_i^{(l)}\hat{\Phi} - \hat{\Phi} \hat{L}_i^{(n)}, \\
&&\hat{Y}_i\hat{\Phi} = \hat{S}_i^{(l)}\hat{\Phi} - \hat{\Phi} \hat{S}_i^{(n)}, \end{array}\right\} \quad \quad \hat{\Phi}\in \mathcal{H}_{nl}.
\label{su2_bimodule}
\end{eqnarray}

The adjoint operators $\hat{W}_i$ and $\hat{Y}_i$ also generate a $SU(2) \star SU(2)$. 
The  $SU(2)$ generated by $\hat{W}_i$  is a reducible  representation which is a direct product of two irreducible representations $\frac{l}{2}\otimes \frac{n}{2}$. Similarly the  $SU(2)$ generated by $\hat{Y}_i$ also corresponds to the direct product $\frac{l}{2}\otimes \frac{n}{2}$.
Thus the elements of  $\mathcal{H}_{nl}$ can be expanded in terms of the eigenfunctions of $\hat{W}^2$, $\hat{W}_3$ and $\hat{Y}_3$ belonging to the irreducible decomposition of $\frac{l}{2}\otimes \frac{n}{2}$ of each $SU(2)$:
\begin{equation}
\frac{l}{2}\otimes \frac{n}{2}= \frac{|l-n|}{2} \oplus \left(\frac{|l-n|}{2}+1\right) \oplus \left(\frac{|l-n|}{2}+2\right) \oplus...\oplus\frac{l+n}{2}.
\end{equation}
$\mathcal{H}_{nl}$ is spanned by the operators $\hat{\Phi}^{j}_{J,\kappa,m}$ which satisfy
\begin{eqnarray}\left.\begin{array}{lll}
&&\hat{W}^2\hat{\Phi}^{jm^\prime}_{J,\kappa,m} =j(j+1) \hat{\Phi}^{jm^\prime}_{J,\kappa,m}, \\
&&\hat{W}_3\hat{\Phi}^{jm^\prime}_{J,\kappa,m} =m \hat{\Phi}^{jm^\prime}_{J,\kappa,m}, \\
&&\hat{Y}_3\hat{\Phi}^{jm^\prime}_{J,\kappa,m} = m^\prime \hat{\Phi}^{jm^\prime}_{J,\kappa,m}, 
\end{array}\right\} 
\quad\quad -j \leq m,m^\prime \leq j.
\end{eqnarray}
We will specify the allowed  values of $j$ in the following discussion.

The operator $\hat{f} = \tilde{N}_{\tilde{l}\tilde{n}}(\hat{a}_1^\dagger)^{\tilde{l} } (\hat{a}_4)^{\tilde{n}}$ (where $\tilde{N}_{\tilde{l}\tilde{n}}$ is a constant)  is an element of $\mathcal{H}_{nl}$ if $0\leq \tilde{n}\leq n$ and $\tilde{l}-\tilde{n}=l-n=\kappa$.
It is easy to see that the operator $\hat{f}$ satisfies
\begin{eqnarray}
&&\hat{W}_+ \hat{f} = [\hat{a}_1^\dagger \hat{a}_2+\hat{a}_3^\dagger \hat{a}_4, \hat{f}] = 0, \nonumber \\
&&\hat{Y}_+ \hat{f} = [\hat{a}_1^\dagger \hat{a}_3+\hat{a}_2^\dagger \hat{a}_4, \hat{f}] = 0, \nonumber \\
&&\hat{W}_3\hat{f} =\frac{1}{2} [\hat{a}_1^\dagger \hat{a}_1+\hat{a}_3^\dagger \hat{a}_3-\hat{a}_2^\dagger \hat{a}_2-\hat{a}_4^\dagger \hat{a}_4, \hat{f}] = \frac{\tilde{l}+\tilde{n}}{2} \hat{f},\\
&&\hat{Y}_3 \hat{f} = \frac{1}{2} [\hat{a}_1^\dagger \hat{a}_1+\hat{a}_2^\dagger \hat{a}_2-\hat{a}_3^\dagger \hat{a}_3-\hat{a}_4^\dagger \hat{a}_4, \hat{f}] = \frac{\tilde{l}+\tilde{n}}{2} \hat{f}. \nonumber 
\end{eqnarray}

So $\hat{f}$ is the highest weight vector for both the $SU(2)$'s  in the representation with $j=\frac{\tilde{l}+\tilde{n}}{2}$.  We will denote this highest weight vector by $\hat{\Phi}_{J,\kappa, j}^{j,j}$:
\begin{equation}
\hat{\Phi}_{J,\kappa,j}^{j,j}=N^{j,j}_{J,\kappa,j}(\hat{a}_1^\dagger)^{\tilde{l} } (\hat{a}_1)^{\tilde{n}}, \quad\quad\quad J=\frac{l+n}{2},
\end{equation}
where $N^{j,j}_{J,\kappa,j}$ is a constant.

All the lower weight vectors can be generated by the action of  $W_-$ and $Y_-$ on  $\hat{\Phi}_{J,\kappa,j}^{j,j}$
\begin{eqnarray}
&&(\hat{W}_- )^{j-m} \hat{\Phi}_{J,\kappa,j}^{j,j} = (N^{\prime})^{j,j}_{J,\kappa, m} \hat{\Phi}_{J,\kappa,m}^{j,j}, \nonumber \\
&&(\hat{Y}_-)^{j-m^\prime}\hat{\Phi}_{J,\kappa,j}^{j,j} = (N^{\prime\prime})^{j,m\prime}_{J,\kappa, j} \hat{\Phi}_{J,\kappa,j}^{j, m^\prime},  \\
&&(\hat{W}_-)^{j-m} (\hat{Y}_-)^{j-m^\prime} \hat{\Phi}_{J,\kappa,j}^{j,j}= (N^{\prime\prime\prime})^{j, m^\prime}_{J,\kappa, m} \hat{\Phi}_{J,\kappa,m}^{j, m^\prime}. \nonumber 
\end{eqnarray}

As $\tilde{n}$ takes all integer values from 0 to $n$, $j$ takes values $\frac{\kappa}{2} \leq j \leq J$.
Thus $\mathcal{H}_{nl}$ is spanned by the operators 
\begin{equation}
\hat{\Phi}^{j, m^\prime}_{J,\kappa,m}  \quad\quad \mathrm{with}\quad  -j\leq m, m^\prime\leq j,\quad \quad j=\frac{\kappa}{2}, \frac{\kappa}{2}+1, ...... J.
\end{equation}
The number of linearly independent operators which spans the space $\mathcal{H}_{nl}$ is 
\begin{equation}
d=\sum_{j=\frac{\kappa}{2}}^J \left(2j+1\right)^2,
\end{equation}
which is the dimension of $\mathcal{H}_{nl}$ .

An arbitrary element $\Phi$ of $\mathcal{H}_{nl}$ can be expressed as
\begin{equation}
\hat{\Phi}= \sum_{j=\frac{\kappa}{2}}^J \sum_{m^\prime=-j}^j \sum_{m=-j}^j c^{j m^\prime}_{J,\kappa,m},\hat{\Phi}^{j, m^\prime}_{J,\kappa,m},\quad \quad c^{j m^\prime}_{J,\kappa,m}\in \mathbb{C}.
\end{equation}

The {\it topological charge} operator is defined as
\begin{equation}
\hat{K}_0\equiv\frac{1}{2}[\hat{N},\quad],
\label{fuzzy_topo_charge1}
\end{equation}
which satisfies
\begin{equation}
\left[\hat{W}_i, \hat{K}_0\right]=0=\left[\hat{Y}_i, \hat{K}_0\right],\quad\quad  \left[\hat{W}^2, \hat{K}_0\right]=0.
\end{equation}

An element $\hat{\Phi} \in \mathcal{H}_{nl}$ is also an eigenfunction of $\hat{K}_0$:
\begin{equation}
\hat{K}_0\hat{\Phi}\equiv\frac{1}{2}[\hat{N},\hat{\Phi}] =\frac{\kappa}{2}\hat{\Phi}
\label{fuzzy_topo_charge2}
\end{equation}
where $\kappa$  takes  integer values.
Thus the operators $\hat{\Phi} \in \mathcal{H}_{nl}$ are identified as the noncommutative analogue of the sections of the complex line bundle with topological charge $\kappa$.

\subsection{Illustrations}

Let us illustrate the above result in some simple cases.
Consider the space  of linear operators $\mathcal{H}_{12}$  which map $\tilde{\mathcal{F}}_1 \rightarrow \tilde{\mathcal{F}}_2$.
In this case,
\begin{equation}
 n=1,\quad\quad l=2,\quad \quad \kappa=|l-n|=1,\quad \quad J=\frac{l+n}{2}=\frac{3}{2}.
\end{equation}
So $j$ takes  values $\frac{1}{2}$ and $\frac{3}{2}$.

The space $\mathcal{H}_{12}$ is spanned by operators which are generated by the technique described in the previous subsection:
\begin{equation}j=\frac{1}{2}:\quad \left\{\quad\quad\quad
 \begin{CD} \hat{a}_1^\dagger  @ >{\hat{W}_-}>> \hat{a}_2^\dagger\\
@V{\hat{Y}_-}VV @VV{\hat{Y}_-}V\\
\hat{a}_3^\dagger@>>{\hat{W}_-}> \hat{a}_4^\dagger \end{CD} 
\right.
\label{j_half_representation}
\end{equation}
\begin{equation}j=\frac{3}{2}:\quad \left\{\quad\quad\quad
\begin{CD} p_{11}  @ >{\hat{W}_-}>> p_{12}@ >{\hat{W}_-}>> p_{13}@ >{\hat{W}_-}>> p_{14}\\
@V{\hat{Y}_-}VV @VV{\hat{Y}_-}V@V{\hat{Y}_-}VV @VV{\hat{Y}_-}V\\
p_{21} @ >{\hat{W}_-}>> p_{22}@ >{\hat{W}_-}>> p_{23}@ >{\hat{W}_-}>> p_{24} \\
@V{\hat{Y}_-}VV @VV{\hat{Y}_-}V@V{\hat{Y}_-}VV @VV{\hat{Y}_-}V\\
p_{31}@>{\hat{W}_-}>> p_{32}@ >{\hat{W}_-}>> p_{33}@ >{\hat{W}_-}>> p_{34} \\
@V{\hat{Y}_-}VV @VV{\hat{Y}_-}V@V{\hat{Y}_-}VV @VV{\hat{Y}_-}V\\
p_{41}@>{\hat{W}_-}>> p_{42}@ >{\hat{W}_-}>> p_{43}@ >{\hat{W}_-}>> p_{44} \end{CD} 
\right.
\label{j_3/2_representation}
\end{equation}
where the various operators in the grid are
\begin{eqnarray}\left. \begin{array}{lll}
&&p_{11}=(\hat{a}_1^\dagger)^2 \hat{a}_4,\\
&& p_{12}=\hat{a}_1^\dagger\hat{a}_2^\dagger \hat{a}_4 - (\hat{a}_1^\dagger)^2 \hat{a}_3, \\
&&... \\
&&...\\
&&p_{43}=(\hat{a}_4^\dagger)^2\hat{a}_2- \hat{a}_3^\dagger\hat{a}_4^\dagger\hat{a}_1, \\ 
&& p_{44}=(\hat{a}_4^\dagger)^2\hat{a}_1.\end{array} \right\}
\end{eqnarray}
For simplicity of presentation, we have omitted the normalizations in the definition of the vectors in the above diagram.

\section{$T^{p,q}$, $T^{p,0}$  and Their  Fuzzy Analogues }\label{tpq}
There exist a class of manifolds $T^{p,q}$ which can be the base of the conifold $Y^6$ \cite{Candelas:1989js}. $X^5$ is just one such manifold.  
All $T^{p,q}$'s  are topologically $S^3 \times S^2$, but they are geometrically inequivalent.  We have already dealt with $X^5(=T^{1,1})$ 
in detail earlier. Here we discuss $T^{p,q}$ in general and compare the different geometries of the base.

The metric on $T^{p,q}$ is given by 
\begin{eqnarray}
ds^2_{pq}=\frac{\rho^2}{c^2_1}(d\phi+  p\cos \theta_1 d \zeta_1 + q\cos \theta_2 d \zeta_2)^2+ \frac{\rho^2}{c^2_2} (d\theta_1^2 + \sin ^2 \theta_1 d\zeta_1^2)\nonumber\\ +\frac{\rho^2}{c^3_3} (d\theta_2^2 + \sin ^2 \theta_2 d\zeta_2^2),\label{metric_n_pq}\\
0\leq\theta_a< \pi,\quad \quad 0\leq\zeta_a< 2\pi, \quad \quad 0\leq\phi< 4\pi,\nonumber
\end{eqnarray}
where $c_1, c_2, c_3, p$ and $q$ are constants (their values are given later).
Demanding  Ricci-flatness of  the metric on  $Y^6$ 
\begin{equation}
ds^2_{Y^6} = d\rho^2 +ds^2_{pq}
\label{metric_cone21}
\end{equation}
leads to
\begin{equation}
\tilde{R}_{ab}= \frac{4}{\rho^2} \tilde{g}_{ab}.
\label{einstein_cond3}
\end{equation}
The Einstein condition (\ref{einstein_cond3}) enforces that  $c_1, c_2, c_3, p$ and $q$ must satisfy
\begin{eqnarray}
p=\frac{\sqrt{2(c_2^2-4)}}{c_2^2}c_1,\nonumber\\
q=\frac{\sqrt{2(c_3^2-4)}}{c_3^2}c_1,\\
c_2^2+c_3^2-12=0.\nonumber
\end{eqnarray}

The manifolds $T^{p,q}$ are $U(1)$ fibre bundles over $S^2 \times S^2$ for integer values of $p$ and $q$.
The $S^2 \times S^2$ is given by 
\begin{equation}\left.
\begin{array}{lll}
y_i =y_i (\theta_1,\zeta_1), \quad\quad  y_i y_i =\textrm{fixed}, \\ \\
w_i =w_i (\theta_2,\zeta_2), \quad\quad  w_i w_i =\textrm{fixed},
\end{array}\right\}\quad \quad \quad i=1,2,3.
\end{equation}
In analogy with (\ref{topo_charge_1}), there exist differential operators $K_0^{(y)}$ and $K_0^{(w)}$ on the spheres $S^2$ generated by $y_i$ and $w_i$ respectively. The integers $p$ and $q$ are related to the eigenvalues of $K_0^{(y)}$ and $K_0^{(w)}$ respectively and correspond to the {\it Chern numbers} of the principal bundle specified by  $(p,q)$.

For $p=q=1$, we get $c_2^2=c^2_3=6$, $c_1^2=9$ and the metric (\ref{metric}) on $X^5$ ($= T^{1,1}$).
In this case, 
\begin{equation}
K_0^{(w)}\equiv K_0^{(y)}\equiv K_0
\end{equation}
where $K_0$ is given in (\ref{topo_charge_1}). This is a  principal bundle with Chern numbers $(1,1)$.  The line bundles associated with this principal bundle have charge $(\kappa,\kappa)$.

For  $p=1, q=0$, we get $
c_3^2=4$, $c_2^2=8$, $c_1^2=8$
and the metric becomes
\begin{equation}
ds^2_{10}
=\frac{\rho^2}{8}{\Big[}(d\phi+  \cos \theta_1 d \zeta_1 )^2+  (d\theta_1^2 + \sin ^2 \theta_1 d\zeta_1^2){\Big]} +\frac{\rho^2}{4} {\Big(}d\theta_2^2 + \sin ^2 \theta_2 d\zeta_2^2{\Big)}.
\label{metric_s3_s2_1}
\end{equation}
The first term in (\ref{metric_s3_s2_1}) is the metric on $S^3$ of radius $R= \frac{\rho}{\sqrt{2}}$ and
$(\phi, \theta_1, \zeta_1)$ are the Euler angles on $S^3$. The second term is the metric on $S^2$ of radius $r= \frac{\rho}{2}$.
So  (\ref{metric_s3_s2_1}) is the metric on $S^3 \times S^2 (= T^{1,0})$.

If we set $R=1$, we get
\begin{eqnarray}
ds^2_{10}=\frac{1}{4}{\Big(}d\theta_1^2 + d\zeta_1^2 + 2 \cos\theta_1 d\zeta_1d\phi+d\phi^2 {\Big)}+\frac{1}{2} {\Big(}d\theta_2^2 + \sin ^2 \theta_2 d\zeta_2^2{\Big)}.
\end{eqnarray}

Under the coordinate re-labeling
\begin{eqnarray}
&&\zeta_a \rightarrow \xi_a = 2\pi -\zeta_a, \quad \quad \phi \rightarrow 2\psi=\phi-\xi_1,
\end{eqnarray}
the above metric can be rewritten as
\begin{eqnarray}
ds^2_{10}={\Big(}d\psi+\sin^2\frac{\theta_1}{2} d \xi_1 {\Big)}^2+ {\Big(}\frac{d\theta_1^2}{4} + \frac{1}{4}\sin ^2 \theta_1 d\xi_1^2{\Big)}+\frac{1}{2} {\Big(}d\theta_2^2 + \sin ^2 \theta_2 d\xi_2^2{\Big)}.
\label{metric_s3_s2_2}
\end{eqnarray}

Though both $T^{1,0}$ and $T^{1,1}$ are topologically $S^3\times S^2$, they  are different geometries. $S^3 \times S^2=SU(2) \times \frac{SU(2)}{U(1)}$,  whereas for $X^5$,  the $U(1)$ is quotiented ``democratically" from  each of the $SU(2)$'s:
\begin{eqnarray}
X^5\simeq\frac{SU(2) \times SU(2)}{U(1)}.
\end{eqnarray}
We can compare the volumes of these two spaces: 
\begin{eqnarray}
(Volume)_{11} =\frac{16\pi^3}{27}\rho^5,\quad
(Volume)_{10}=\frac{\pi^3}{\sqrt{2}}\rho^5.
\end{eqnarray}
Thus there is no diffeo which takes us from one to the other. Moreover, the metric (\ref{metric}) is compatible with the K\"{a}hler structure  of the cone $Y^6$ but the metric (\ref{metric_s3_s2_1}) is not. Nevertheless, both $X^5$ and  $S^3\times S^2$ are $U(1)$ fibre bundles over $S^2 \times S^2$.

We will describe the above principal bundle using Hopf fibration as their fuzzification is transparent in this language. We have already discussed $T^{1,1}$ in the previous sections. Below we discuss $T^{1,0}$.

Embedded in $\mathbb{C}^4$, the $S^3 \times S^2$ is described by 
\begin{eqnarray}\left.\begin{array}{lll}
&&\eta_a = z_a, \\ \\
&&q_i = \frac{1}{\sqrt{2}} \bar{z}_c (\sigma_i)_{cd}z_d , \end{array}\right\}\quad \quad a=1,2,\quad c,d=3,4,\quad i=1,2,3
\end{eqnarray}
with
\begin{eqnarray}
\bar{\eta}_a \eta_a =1, \quad\quad q_i q_i =\frac{1}{2}.
\end{eqnarray}
The fibration from $S^3 \times S^2$ to $S^2 \times S^2$ is given by 
\begin{eqnarray}
&&y_i = \frac{1}{2} \bar{\eta}_a (\sigma_i)_{ab}\eta_b, \,\,\,\quad\quad \quad a,b=1,2,\nonumber\\
&& w_i =q_i,\quad\quad\quad\quad\quad\quad\quad\quad i=1,2,3,
\end{eqnarray}
with
\begin{eqnarray}
 y_iy_i=\frac{1}{4}, \quad\quad w_i w_i =\frac{1}{2}.
\end{eqnarray}

The $S^2  \times S^2$ is  parametrized by  angular variables in the metric (\ref{metric_s3_s2_1}): $(\theta_1,\xi_1)$ describes the $S^2$ generated by $y_i$ while $(\theta_2,\xi_2)$ describes the $S^2$ generated by $w_i$.  The $U(1)$ fibre is described by the angular coordinate $\psi$.

The metric (\ref{metric_s3_s2_2}) is of the Kaluza-Klein form (\ref{kk_metric}) with the dilaton $g$ set 1. The metric on the base $S^2 \times S^2$ 
\begin{eqnarray}
g_{\mu\nu}=\left(
\begin{array}{cccc}
\frac{1}{4}&0&0&0\\
0& \frac{1}{4}\sin^2 \theta_1&0&0\\
0&0&\frac{1}{2}&0\\
0&0&0&\frac{1}{2}\sin^2 \theta_2
\end{array}\right)
\label{metric_s2_25_s2_5}
\end{eqnarray}
 is not an Einstein metric
but the metric (\ref{metric_s3_s2_2}) is Einstein.

The gauge fields in this case  are 
\begin{eqnarray}
  &&A^{NN}_{\theta_1} = A^{NS}_{\theta_1} = 0,\quad \quad A^{NN}_{\xi_1} = A^{NS}_{\xi_1} = \frac{\kappa}{2}(1-\cos \theta_1), \nonumber\\
 &&A^{NN}_{\theta_2} = A^{NS}_{\theta_2} = 0, \quad\quad A^{NN}_{\xi_2} = A^{NS}_{\xi_2} =0  ,\nonumber\\ \\ 
 &&A^{SN}_{\theta_1} = A^{SS}_{\theta_1} = 0, \quad\quad A^{SN}_{\xi_1} = A^{SS}_{\xi_1} = -\frac{\kappa}{2}(1+\cos \theta_1), \nonumber\\
&& A^{SN}_{\theta_2} = A^{SS}_{\theta_2}=0,\quad \quad A^{SN}_{\xi_2} = A^{SS}_{\xi_2} = 0  = 0.\nonumber
\end{eqnarray}

The charge operators in this case are
\begin{eqnarray}\left.\begin{array}{ccc}
K_0^{(y)}=\frac{1}{2}\left(\bar{z}_a\frac{\partial}{\partial \bar{z}_a} - z_a\frac{\partial}{\partial z_a}\right),\\ 
\\
K_0^{(w)}=\frac{1}{2}\left(\bar{z}_c\frac{\partial}{\partial \bar{z}_c} - z_c\frac{\partial}{\partial z_c}\right),
\end{array}\right\}  \quad \quad \quad \quad\quad a=1,2, \quad c=3,4.
\end{eqnarray}
The principal bundle has  charge (1,0) and the higher charge bundles have charge $(\kappa ,0)$.

\subsection{ \bf $S_F^3 \times S_F^2 \rightarrow S_F^2 \times S_F^2$}\label{tpq1}

In the fuzzy case too, we can get $S^2_F \times S^2_F$ from both $X^5_F$ (or $T^{1,1}_F$) and $S^3_F \times S^2_F$ (or $T^{1,0}_F$).
We have already described the fuzzy map $X^5_F \rightarrow S^2_F \times S^2_F$ in section \ref{fuzzy_fibre_bundle}. Here let us briefly describe the map  $S^3_F\times S^2_F \rightarrow S^2_F \times S^2_F$.

In  $\mathbb{C}_F^4$ let us define the operators
\begin{eqnarray}
&&\hat{\chi}_1 = \hat{a}_1 \frac{1}{\sqrt{\hat{N}_1+\hat{N}_2}}, \quad \hat{\chi}_2 = \hat{a}_2  \frac{1}{\sqrt{\hat{N}_1+\hat{N}_2}}, \\
&&\hat{\eta}_1 = \hat{a}_3 \frac{1}{\sqrt{\hat{N}_3+\hat{N}_4}}, \quad \hat{\eta}_2 = \hat{a}_4  \frac{1}{\sqrt{\hat{N}_3+\hat{N}_4}}
\end{eqnarray}
and the map 
\begin{eqnarray}
\hat{\Omega}_b= \hat{\chi}_b, \quad \quad \hat{q}_i= \frac{1}{\sqrt{2}} \hat{\eta}_b^\dagger (\sigma_i)_{bc} \hat{\eta}_c, \quad\quad i=1,2,3, \quad\quad b,c=1,2.
\end{eqnarray}
It is easy to see that  
\begin{eqnarray}
\hat{\Omega}_b^\dagger \hat{\Omega}_b =1, \quad\quad \hat{q}_i \hat{q}_i =\frac{1}{\sqrt{2}}\left( \frac{1}{\sqrt{2}}+\frac{1}{\hat{N}_3+\hat{N}_4}\right).
\end{eqnarray}

 Let $\mathcal{G}_{s}$ be the subspace of the Fock space $\mathcal{F}$ defined as
\begin{eqnarray}
\mathcal{G}_{s}= span\{|n_1,n_2,n_3,n_4\rangle: \,\,\, n_3+n_4=s=\textrm{fixed}\}, \quad\quad \mathcal{G}_{s}\subset\mathcal{F}.
\end{eqnarray}

In $\mathcal{G}_{s}$, $\hat{q}_i\hat{q}_i$ takes a fixed value $\frac{1}{\sqrt{2}}\left(\frac{1}{\sqrt{2}}+\frac{1}{s}\right)$. Hence the restriction of the action of the operators $\hat{a}_\alpha$ to the subspace $\mathcal{G}_{s}$ describes the fuzzy space $S^3_F \times S^2_F$.

We define the operator map
\begin{eqnarray}\left.\begin{array}{lll}
\hat{w}_i = \hat{q}_i,\\\\
\hat{y}_i = \frac{1}{2} \hat{\Omega}_b^\dagger (\sigma_i)_{bc} \hat{\Omega}_c, 
\end{array}\right\}
\quad\quad\quad i=1,2,3, \quad\quad b,c=1,2.
\end{eqnarray}
The operators $\hat{w}_i$ and $\hat{y}_i$ satisfy 
\begin{eqnarray}
&&\left[\hat{w}_i, \hat{N}_3+\hat{N}_4\right]=0, \nonumber\\ 
&&\left[\hat{w}_i, \hat{w}_j\right]=\frac{i}{\sqrt{2}}\epsilon_{ijk} \frac{1}{\hat{N}_3+\hat{N}_4}\hat{w}_k, \nonumber \\
&&\left[\hat{y}_i, \hat{N}_1+\hat{N}_2\right]=0, \\
&&\left[\hat{y}_i, \hat{y}_j\right]=i\epsilon_{ijk}\frac{1}{\hat{N}_1+\hat{N}_2} \hat{y}_k, \nonumber\\
&& \left[\hat{w}_i, \hat{y}_j\right]=0 \nonumber
\end{eqnarray}
and it is easy to show that
\begin{equation}
\hat{w}_i \hat{w}_i = \frac{1}{\sqrt{2}}\left( \frac{1}{\sqrt{2}}+\frac{1}{\hat{N}_3+\hat{N}_4}\right),\quad\quad \hat{y}_i \hat{y}_i = \frac{1}{2}\left(\frac{1}{2}+\frac{1}{\hat{N}_1+\hat{N}_2}\right).
\end{equation}

$\mathcal{G}_{s}^n$ is a  subspace of $\mathcal{G}_{s}$ defined as
\begin{eqnarray}
\mathcal{G}_{s}^n= span\{|n_1,n_2,n_3,n_4\rangle: \,\,\,  n_1+n_2=n=\textrm{fixed},\,\,\,n_3+n_4=s=\textrm{fixed}\}
\end{eqnarray}
and 
\begin{equation}
\mathcal{G}_{s}=\oplus_n\mathcal{G}_{s}^n.
\end{equation}

In the space $\mathcal{G}_{s}^n$, 
\begin{eqnarray}
&&\hat{w}_i \hat{w}_i = \frac{1}{\sqrt{2}}\left(\frac{1}{\sqrt{2}}+\frac{1}{s}\right)=\textrm{fixed},\\
&&\hat{y}_i \hat{y}_i = \frac{1}{2}\left(\frac{1}{2}+\frac{1}{n}\right)=\textrm{fixed}.
\end{eqnarray}
So the restriction of the action of the operators $\hat{a}_\alpha$ to the subspace $\mathcal{G}_{s}^n$ describes the fuzzy space $S^2_F \times S^2_F$.

The set of operators
\begin{equation}
\hat{S}_i=\left(\hat{N}_1+\hat{N}_2\right)\hat{y}_i 
\end{equation}
generate the $SU(2)$ algebra
\begin{eqnarray}
\left[\hat{S}_i, \hat{S}_j\right]=i\epsilon_{ijk} \hat{S}_k.
\end{eqnarray}
The space $\mathcal{G}_{s}^n$ is the carrier space of  $(n+1)$-dimensional representation of the $SU(2)$ generated by $\hat{S}_i$'s.

Next we proceed in a similar way as we did for the case of $X^5_F$. $\mathcal{H}^{s}_{nl}$ is the space of operators maps from $\mathcal{G}_{s}^n \rightarrow\mathcal{G}_{s}^l$. In this case, the elements of $\mathcal{H}^{s}_{nl}$ should be identified as the sections of the noncommutative vector bundle.  

An element $\hat{\Psi}$ of $\mathcal{H}^{s}_{nl}$ 
\begin{eqnarray}
&&\hat{\Psi}: \mathcal{G}_{s}^n  \rightarrow\mathcal{G}_{s}^l,\quad \quad \hat{\Psi} \in \mathcal{H}^{s}_{nl}
\end{eqnarray}
is a rectangular matrix of dimension $(l+1) \times (n+1)$. This space is spanned by basis vectors
\begin{equation}
\left(\hat{\Omega}^\dagger_1 \right)^{\tilde{l}_1} \left(\hat{\Omega}_2^{\dagger}\right)^{\tilde{l}_2}\hat{\Omega}_1^{\tilde{n}_1} \hat{\Omega}_2^{\tilde{n}_2}
\end{equation}
\begin{equation}
\textrm{such that} \left\{\begin{array}{lll}
\tilde{l}_1+\tilde{l}_2-\tilde{n}_1-\tilde{n}_2 = l-n, \\
\tilde{n}_1+\tilde{n}_2 \leq n.
\end{array}\right.
\label{condition}
\end{equation}
So an arbitrary element $\hat{\Psi}$ can be expressed as
\begin{eqnarray}
&&\hat{\Psi}= \sum_{\tilde{l}_1, \tilde{l}_2,\tilde{n}_1, \tilde{n}_2} c_{\tilde{l}_1 \tilde{l}_2\tilde{n}_1 \tilde{n}_2}  \left(\hat{\Omega}^\dagger_1 \right)^{\tilde{l}_1} \left(\hat{\Omega}_2^{\dagger}\right)^{\tilde{l}_2}\hat{\Omega}_1^{\tilde{n}_1} \hat{\Omega}_2^{\tilde{n}_2}
\end{eqnarray}
with $\tilde{l}_1,\tilde{l}_2, \tilde{n}_1$ and $\tilde{n}_2$ satisfying the conditions (\ref{condition}).

Just as we did in section \ref{fuzzy_fibre_bundle}, we can express $\hat{\Psi}$ in the basis of eigenfunctions of $Ad (\hat{S}_3)$ and $Ad (\hat{S}_i \hat{S}_i)$.

The topological charge operator on the fuzzy spheres generated by $\hat{y}_i$ and $\hat{w}_i$ are given by
\begin{eqnarray}\left.\begin{array}{lll}
&&\hat{K}^{(\hat{y_i})}_0 =\frac{1}{2} \left[\hat{N}_1+\hat{N}_2, \,\,\,\right], \\ \\
&&\hat{K}^{(\hat{w_i})}_0 =\frac{1}{2} \left[\hat{N}_3+\hat{N}_4, \,\,\,\right]. \\
\end{array} \right.
\end{eqnarray}

The operators $\hat{\Psi}$ satisfy 
\begin{equation}
\hat{K}^{(\hat{y_i})}_0\hat{\Psi} =\frac{\kappa}{2} \hat{\Psi} ,\quad\quad
\hat{K}^{(\hat{w_i})}_0 \hat{\Psi}  =0.
\end{equation}
Using the convention as in the commutative case, we can denote the fuzzy fibre bundle by the pair $(\kappa,0)$.

In case of $X^5_F$ or $T^{1,1}_F$, 
\begin{equation}
\hat{K}^{(\hat{y_i})}_0 =\hat{K}^{(\hat{w_i})}_0=\hat{K}_0
\end{equation}
where $\hat{K}_0$ is given by (\ref{fuzzy_topo_charge1}). In this case the fuzzy bundle is characterized by the pair $(\kappa,\kappa)$.

\end{document}